\documentclass[journal]{IEEEtran}
\usepackage{graphicx}

\usepackage{subcaption}
\usepackage{multirow} 
\usepackage{booktabs,caption,fixltx2e}
\usepackage[flushleft]{threeparttable}
\ifCLASSINFOpdf
\else
\fi
\begin{document}
%
\title{High Frequency Remote Monitoring of Parkinson's Disease via Smartphone: Platform Overview and Medication Response Detection}
%
%
%

\author{Andong~Zhan,
				Max~A.~Little,
				Denzil~A.~Harris,
				Solomon~O.~Abiola,
				E.~Ray~Dorsey,
				Suchi~Saria,
				and Andreas~Terzis%
\thanks{This work was supported in part by Andreas Terzis.}
\thanks{A. Zhan, S. Saria, and A. Terzis are with the Department
of Computer Science, Johns Hopkins University, Baltimore, MD, USA.}
\thanks{D. A. Harris, S. O. Abiola, and E. R. Dorsey are members of Center for Human Experimental Therapeutics, University of Rochester Medical center, Rochester, NY, USA; E. R. Dorsey is also with the Department of Neurology, University of Rochester Medical center, Rochester, NY, USA}
\thanks{M. A. Little is with the Nonlinearity and Complexity Research Group, Aston University, Birmingham, UK; he is also with the Media Lab, Massachusetts Institute
of Technology, Cambridge, MA, USA.}%
}

\maketitle

\begin{abstract}
\textit{Objective}: 
The aim of this study is to develop a smartphone-based high-frequency
remote monitoring platform, assess its feasibility for remote
monitoring of symptoms in Parkinson¡¯s disease, and demonstrate
the value of data collected using the platform by detecting dopaminergic
medication response.

\textit{Methods}: We have developed HopkinsPD, a novel smartphone-based monitoring platform, which measures symptoms actively (i.e. data are collected when a suite of tests is initiated by the individual at specific times during the day), and passively (i.e. data are collected continuously in the background). After data collection, we extract
features to assess measures of five key behaviors related to PD symptoms --  voice, balance, gait, dexterity, and reaction time. A random forest classifier is used to discriminate measurements taken after a dose of medication (treatment) versus before the medication dose (baseline).

\textit{Results}: A worldwide study for remote PD monitoring was established using HopkinsPD in July, 2014. This study used entirely remote, online recruitment and installation, demonstrating highly cost-effective scalability.
In six months, 226 individuals (121 PD and 105 controls) contributed over 46,000 hours of passive monitoring data and approximately 8,000 instances of structured tests of voice, balance, gait, reaction, and dexterity. To the best of our knowledge, this is the first study to have collected data at such a scale for remote PD monitoring. Moreover, we demonstrate the initial ability to discriminate treatment from baseline with 71.0($\pm$0.4)\% accuracy, which suggests medication response can be monitored remotely via smartphone-based measures.

\end{abstract}


%
\IEEEpeerreviewmaketitle

\section{Introduction}
Parkinson disease (PD) is a progressive neurodegenerative disease associated with substantial morbidity, increased mortality, and particularly high economic burden \cite{weintraub2008parkinson}. The prevalence of PD is increasing with as many as one million Americans and an estimated seven to ten million people worldwide living with PD \cite{pdf_stats}. Direct and indirect costs of Parkinson's are estimated to be nearly \$25 billion annually in the United States alone, and are expected to grow significantly as the number of affected individuals increases \cite{pdf_stats}.

Currently, individuals living with PD have access to care
or participate in research primarily during in-person clinic
or research visits, which take place at most once every few
months. More frequent clinic visits are limited by travel distance,
increasing disability, and uneven distribution of doctors
\cite{achey2014past}. During clinic visits, clinicians assess current disease status
and adjust medication for their patients based on response and
side effects. However, a key challenge for PD treatment is that
PD progression varies among individuals \cite{pdf_progression}. 
Furthermore, individuals may exhibit large variations \cite{puk}. 
More specifically,
symptoms can fluctuate substantially over the medium
term, and the progression is not smooth for everyone \cite{little2013quantifying}.
Accordingly, it is difficult for clinicians to provide optimal
treatment for their patients based on these periodic ``snapshots''
of the disease progression. Therefore, assessments based on
clinic visits alone are insufficient, and high frequency remote
monitoring is needed to improve the quantity and quality of
care for PD. For example, real-time, objective monitoring of
daily fluctuations in symptoms can enable timely assessments
of disease and response to treatment. These data can enable
more subtle adjustments of medications or other therapies for
PD and assessment of the efficacy of novel interventions.

Existing studies have required the use of specialized
and expensive medical devices such as wearable accelerometers
and gyroscopes, EEG and passive infrared sensors \cite{patel2009monitoring}\cite{sejdic2014comprehensive}\cite{tsipouras2011automated}\cite{pavel2007continuous}\cite{sanders2013remote}\cite{begg2005support}\cite{levy2000high}. Many of these studies have also only reported data collected in the laboratory setting \cite{patel2009monitoring}\cite{sejdic2014comprehensive}\cite{tsipouras2011automated}\cite{begg2005support}\cite{levy2000high}, which does not faithfully represent the patterns of variability that individuals with PD may experience at home. In addition, the majority of past studies have focused on monitoring only one aspect of PD such as dyskinesia \cite{tsipouras2011automated}, gait \cite{pavel2007continuous}\cite{arora2014high}\cite{mazilu2012online}\cite{begg2005support}\cite{weiss2011toward}, voice \cite{tsanas2012novel}, postural sway \cite{arora2014high}, and resting tremor \cite{lemoyne2010implementation}\cite{levy2000high}. However, PD is a multi-faceted disease
with many varied symptoms. Thus, a multi-dimensional
approach is needed to continuously monitor all PD symptoms
at home.

Mobile phone based tracking and measurement tools offer a
promising new avenue for monitoring progressive conditions
outside the clinic. The smartphone is becoming one of the
most basic necessities. From a recent survey of 170,000
adult Internet users across 32 markets, 80\% now own a
smartphone \cite{gwi}. Moreover, without the need for expensive specialized medical hardware, new software tools can be easily downloaded and installed on an individual's smartphone for in-home monitoring.

Therefore, we hypothesize that a smartphone based platform
can be used to objectively and remotely measure aspects
related to PD (e.g., voice, balance, dexterity, gait, and reaction
time) and activities of daily living. To validate this hypothesis,
we have 1) developed HopkinsPD, a unified PD-specific
remote monitoring platform that incorporates both active and
passive tests to provide high frequency monitoring of symptoms
and activities of daily living related to PD, 2) tested
the feasibility of an entirely remote, online recruitment and
installation process that enables large-scale, low-cost clinical
studies, and c) demonstrated that a machine learning approach
using the active tests can detect medication response with
71.0($\pm$0.4)\% accuracy.

\section{Background}

The most widely-studied and understood symptoms in PD pertain to impairments in the motor subsystem of the central nervous system, including tremor at rest, bradykinesia, rigidity, and postural instability. Other non-motor aspects of the disease include depression, anxiety, autonomic dysfunction, and dementia, which are common and significantly affect health-related quality of life of both individuals with PD and their caregivers \cite{weintraub2008parkinson}. 

The Unified Parkinson's Disease Rating Scale (UPDRS) is the most widely used tool for assessing symptoms and clinical status over time in PD \cite{weintraub2008parkinson}. The latest revision, MDS-UPDRS \cite{goetz2008movement}, consists of assessments for 1) Non-motor Aspects of Experiences of Daily Living (nM-EDL), 2) Motor Aspects of Experiences of Daily Living (M-EDL), 3) Motor examination, and 4) Motor complications. Currently, UPDRS assessments are obtained subjectively by movement disorder specialists \cite{dorsey2013caring}. 

The rapid rise of wearable consumer devices and smartphone technologies and the need for high frequency monitoring of PD symptoms have led to a proliferation of remote monitoring studies in PD. However, existing studies suffer from the following shortcomings. First, the majority of them rely on specialized medical hardware. For example, studies have used wearable accelerometers and gyroscopes \cite{patel2009monitoring}\cite{sejdic2014comprehensive}\cite{tsipouras2011automated}\cite{weiss2011toward}\cite{mera2012feasibility}, EEG \cite{sanders2013remote}\cite{levy2000high}, treadmill and video camera \cite{begg2005support}, or passive infrared sensors \cite{pavel2007continuous}.  Often these studies require many sensors to be mounted at various positions on the body \cite{patel2009monitoring}\cite{sejdic2014comprehensive}\cite{tsipouras2011automated}. 
Additionally, these commercial medical devices are extremely expensive (often $>$\$3,000 per device excluding software) in comparison to the essential embedded sensing hardware (e.g. MEMS accelerometer, $\sim$\$5) and require the use of proprietary analysis algorithms whose internal operation is not available to scientific scrutiny and independent replication.
These requirements significantly limit the use of this technology in the home and community setting, and for large-scale studies of PD symptoms.
Secondly, existing studies have typically focused on monitoring only one or two aspects (e.g., dyskinesia \cite{tsipouras2011automated}, gait \cite{pavel2007continuous}\cite{arora2014high}\cite{mazilu2012online}\cite{begg2005support}\cite{weiss2011toward}, voice \cite{tsanas2012novel}, postural sway \cite{arora2014high}, tremor \cite{lemoyne2010implementation}\cite{levy2000high}\cite{mera2012feasibility}, and bradykinesia \cite{mera2012feasibility}) of PD. Since in PD no single symptom gives a full picture of an individual's disease state, a multi-dimensional approach is needed to monitor PD comprehensively.
Finally, previous studies \cite{patel2009monitoring}\cite{sejdic2014comprehensive}\cite{tsipouras2011automated}\cite{begg2005support}\cite{levy2000high}\cite{weiss2011toward}\cite{mera2012feasibility} have primarily reported data from monitoring individuals with PD in the clinical laboratory setting, which are therefore geographically and temporally restricted. 
 To the best of our knowledge, our study is the first to collect high frequency PD-specific data objectively and remotely, on such a large, worldwide scale.

Currently, there is no cure for Parkinson's disease, but treatment can help to control the symptoms. For instance, antiparkinsonian medicines, like levodopa, can help control motor symptoms by increasing dopamine in brain. For individuals with PD who have good medication response, the symptoms of PD can be substantially controlled. By contrast, those in the advanced stages of PD may suffer periods of ``wearing off'', i.e. the medication ceases to have any effect, and instead may develop troublesome side effects such as levodopa-induced dyskinesias and problems with impulse control.
Since medication response and side effects vary by individual substantially, a personalized medication regime is crucial to maintain quality of life. However, it is difficult for clinicians to determine the optimal regime based on brief moments of observation during clinic visits.
Monitoring medication response remotely and objectively is one crucial idea for producing individually optimized PD medication regimes. Previous studies\cite{weiss2011toward}\cite{mera2012feasibility}\cite{Griffiths2012} have attempted to monitor medication response remotely and objectively, but are limited in terms of trial size and duration. 
Weiss et al. \cite{weiss2011toward} discovered that gait features improved significantly when the medication effect peaked in a study on 22 PD participants in clinic by using a triaxial accelerometer on their lower back.
Mera et al. \cite{mera2012feasibility} deployed a finger-worn motion sensor to 10 PD participants at home for six days. They showed that rest and postural tremor and the average speed and amplitude of hand movements were improved during controlled tasks, and also highlighted the differences in motor response to medication among individuals.
Griffiths et al. \cite{Griffiths2012} used a wrist-worn motion sensor to continuously monitor 34 PD participants for 10 days. They showed the severity of bradykinesia and dyskinesia before and after the change in medication. 
Compared with these studies, our platform does not require specialized medical hardware. Furthermore, our platform allows cost-effective scaling in terms of numbers of participants and study duration (e.g. monitoring more than 200 participants continuously over six months); while previous studies have only been feasible for very small datasets (less than 34 participants) with short duration (less than ten days). More importantly, previous studies have only explored symptom-related features (e.g. amplitude of gait, speed of hand movement) and response to medications. However, they do not explore the discrimination of medication response.

\section{Methods}

\begin{figure}[h]
\centering
  \centering
  \includegraphics[width=.8\linewidth]{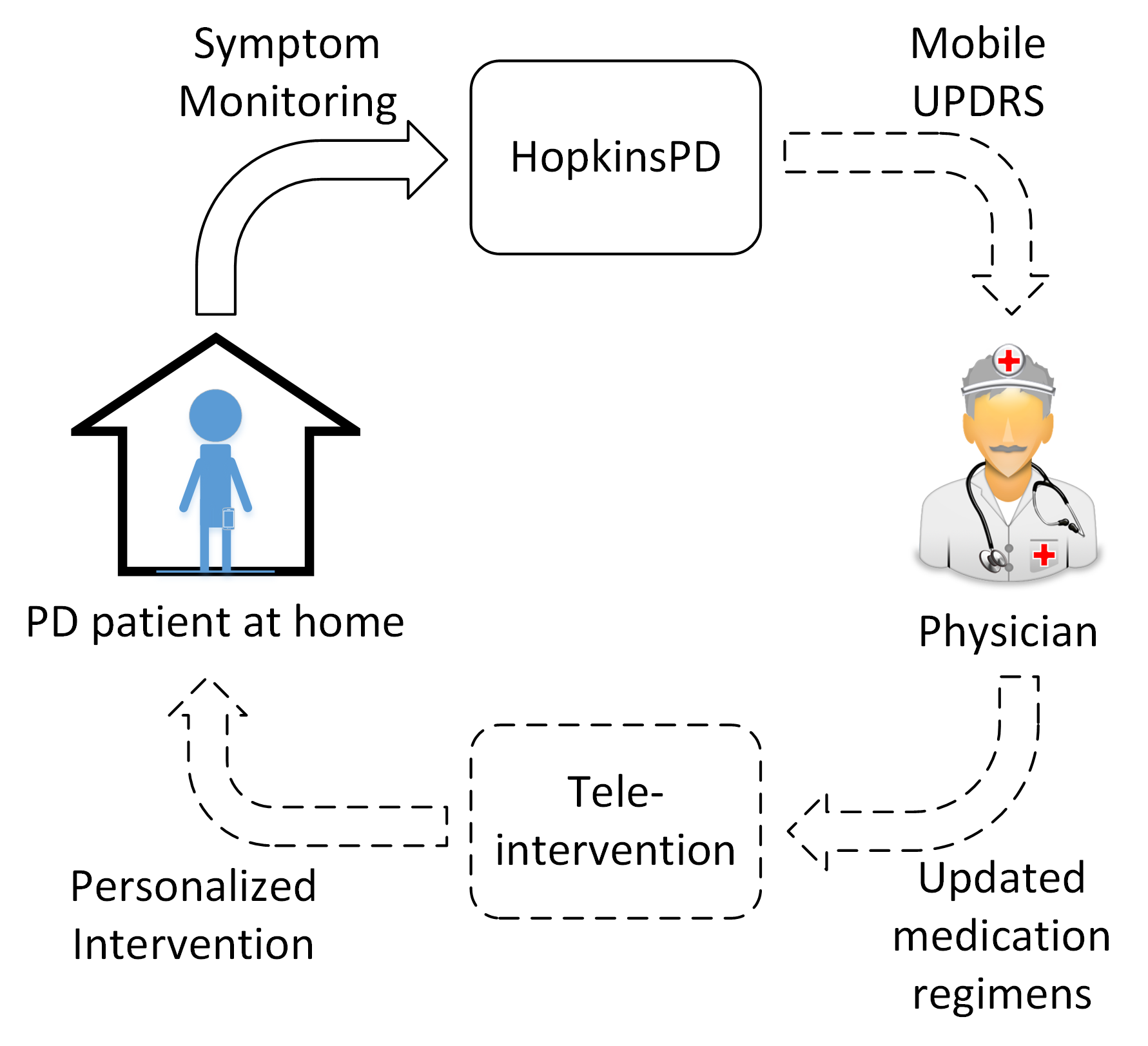}
  \caption{A closed loop for PD monitoring and intervention}
  \label{fig:vision}
\end{figure}
\subsection{HopkinsPD Platform: Architecture and Monitoring Tests Implemented}

\begin{figure*}[t]
  \centering
  \includegraphics[width=.90\linewidth]{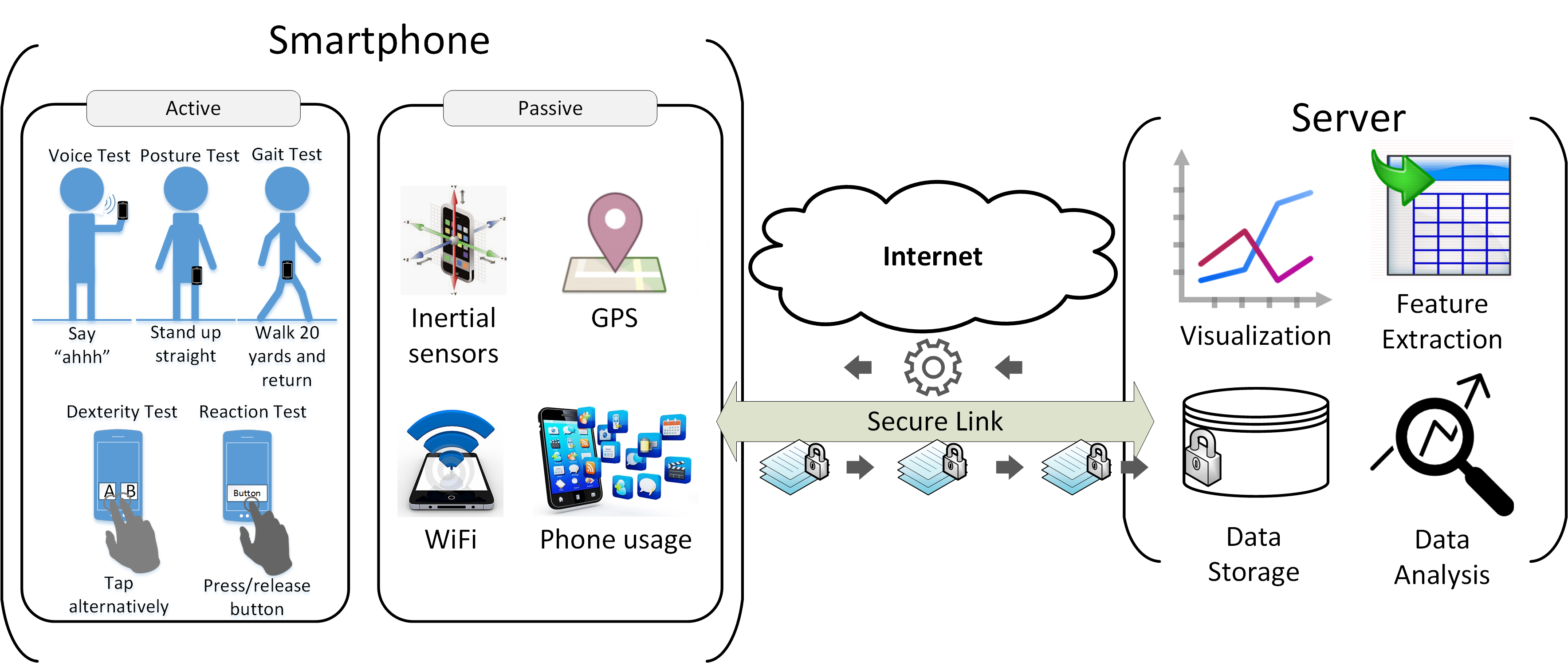}
\caption{HopkinsPD Architecture. See the Supplemental Section for more details on the system.}
\label{fig:arch}
\end{figure*}
\begin{table*}[t]
\footnotesize
\caption{Description of active tests in HopkinsPD}
\centering
	\begin{threeparttable}	
	\centering
	\begin{tabular}{p{2cm} p{8cm} p{2.5cm} p{10em}} 
		\toprule
		\bfseries Test &\bfseries Description provided to participants &\bfseries Relevant PD  &\bfseries Related MDS-UPDRS \\
			& &\bfseries symptoms &\bfseries III Motor exam\\
		\midrule
		Voice & Place the phone to your ear as if making a normal phone call, take a deep breath, and say ``aaah'' for as long and as steadily as you can. & Dysphonia & 3.1 Speech \\
		Balance & Stand up straight unaided and place the phone in your pocket for 30  & Postural instability & 3.12 Postural stability \\
			& seconds.& & 3.13 Posture \\
		Gait & Stand up and place the phone in your pocket. When the buzzer vibrates  & Bradykinesia & 3.8 Leg agility\\
			& walk forward for 20 yards; then turn around and walk back. & Freezing of gait & 3.10 Gait\\
			& & & 3.11 Freezing of gait \\
		Dexterity & Place the phone on a surface such as a desk or table. Tap the buttons & Bradykinesia & 3.4 Finger tapping \\
			& alternately with the index and middle finger of one hand, keeping a  & Reduced dexterity & 3.5 Hand movements\\
			& regular rhythm. & & \\
		Reaction & Keep the phone on a surface as before. Press and hold the on-screen button (i.e., at the bottom of the screen) as soon as it appears; release it as soon as it disappears. & Bradykinesia Reduced dexterity & 3.5 Hand movements \\
		Rest Tremor\tnote{*} & Sit upright, hold the phone in the hand most affected by your tremor, and rest it lightly in your lap. & Resting tremor & 3.17 Rest tremor amplitude \\
		Postural Tremor\tnote{*} & Sit upright and hold the phone in the hand most affected by your tremor outstretched straight in front of you. & Postural Tremor & 3.15 Postural tremor of the hands \\
		\bottomrule
	\end{tabular}
	\begin{tablenotes}	
	\scriptsize
		\item[*]{Tests have been implemented but not used in this study.}
	\end{tablenotes}
	\end{threeparttable}
	\label{tab:tests}
\end{table*}


HopkinsPD allows the monitoring of PD symptoms remotely by using an application installed on the users' own smartphone. The most significant benefit of this mobile-based approach is high accessibility. In particular, it is inexpensive as no additional purchase is needed for smartphone owners and smartphone-based tests can be conducted anywhere and at any time. This enables assessments at high frequency and large scale in terms of numbers of recruits. Furthermore, it can serve as a potential platform for incorporating care, e.g., telemedicine \cite{achey2014past}.
Therefore, HopkinsPD can be seen as a preliminary step towards enabling a complete, ``closed-loop'' remote monitoring and remote intervention tool for PD (Figure \ref{fig:vision}).

HopkinsPD enables fully-automated capture, compression, encryption, and upload to secure server storage by both actively interacting with smartphone users and passively sensing their daily activities (Figure \ref{fig:arch}). The core of the HopkinsPD mobile application is a set of tests to monitor and assess symptoms appearing on the UPDRS scale through smartphones, which consists of:

\begin{enumerate}
	\item \textit{Active tests}, tests that are initiated and self-administered by the participants at various times during the day: these tests are designed to measure several aspects of motor function such as gait, voice, dexterity, reaction time, and postural instability (balance), using built-in smartphone sensors (See more details in Table \ref{tab:tests}, and visual depiction in Figure \ref{fig:app}).
	\item \textit{Passive tests}, running continuously and unobtrusively in the background, are designed to measure aspects of daily living: these use the sensors such as accelerometer, gyroscope, magnetic field strength, GPS location, WiFi parameters, and phone usage logs to measure movement (e.g. whether the individual is experiencing frozen gait or dyskinesia), as well as location and social behavior (e.g. whether they are primarily home-bound or have an active lifestyle). Passive monitoring provides a way to be monitored objectively without interrupting routine activities. The successful monitoring of these daily details may allow comprehensive insight into the behavior and lifestyles of individuals living with PD, which have not been fully investigated in previous studies.
	\item \textit{Self-reported} evaluation of their overall health, mood, and well-being: the significant advantage of such mobile-based questionnaires is that they can be completed outside the clinic. For example, considering that about half to two-thirds of people with PD report that they have memory problems \cite{pdf_cognitive}, an on-demand survey system which can probe such problems may be much more accurate and effective than the current approach, in which questionnaires can only be completed during clinic visits thereby relying on the memory of individuals to accurately report their own symptoms over the last few months.
\end{enumerate}

In addition to data collection, HopkinsPD provides HIPAA-compliant data streaming to a secure server and web-based analysis and visualization of the resulting data. Additional specifics of the HopkinsPD system are summarized in the Supplemental Section.

\subsection{Study Enrollment}
A feasibility study using HopkinsPD was launched in the summer of 2014 to track and quantify characteristics of PD on a worldwide scale  \cite{abiola}. 
In this study, participants were identified and recruited using an email database from the Parkinson’s Voice Initiative, online media, and patient registries such as the Michael J. Fox Foundation’s Fox Trial Finder. They were required to understand English and own an Android smartphone with Internet access (e.g. WiFi).
After enrollment, participants received a confirmation email with an installation URL to click, which automatically installed the application directly onto their smartphone.

During the study, participants were asked to conduct active tests and passive monitoring daily (self-report surveys were not included in this study). 
Each time an individual initiated the active tests within the HopkinsPD application, the application required the user to perform five active tests measuring voice, balance, gait, dexterity, and reaction time sequentially. We refer to these five tests taken by the individual during a single session as an \textit{instance} of active tests. 
The participants were asked to perform two instances of these active tests each day: the first one in the morning just before taking medications, and the second approximately one hour after the first. For healthy controls, they were asked to perform the first in the morning and the second one hour later. 
The characteristics of participants and collected data are summarized in Section \ref{sec:feasibility}.

\subsection{Medication Response Detection by Using Active Tests}
\subsubsection{Feature Extraction}
here, we focus on analyzing the active test data leaving the analysis of passive monitoring data for future work.
Table \ref{tab:feature} provides an exhaustive list of the features extracted from the five types of active tests, along with a brief description of each feature. Acceleration features were based on definitions used in previous studies \cite{arora2013high}\cite{figo2010preprocessing}\cite{arora2015}. Acceleration features were computed from the tri-axial acceleration time series (x, y, and z-axis), as well as the spherical transformation of the tri-axial acceleration time series (i.e., radial distance, polar angle, and azimuth angle). We apply the acceleration features in Table \ref{tab:feature} for these six axes respectively. As the acceleration time series were sampled at irregular time intervals, we applied the Lomb-Scargle periodogram to extract frequency-based features \cite{scargle1982studies} e.g. the dominant frequency component in Hz and its amplitude. All the acceleration features are used by both the balance and gait tests.
To extract the voice features, we first divide the 20-second audio sample into 40 frames leading to 0.5 second frame duration. Then, we tag each frame as containing a `voiced' signal if that frame has amplitude greater than the first quartile of the amplitudes among all frames. Then, for further analysis, we select the longest consecutive run of voiced frames. The length of the largest consecutive run of voiced frames is the ``voice duration'' feature. Other features extracted from these voiced frames include dominant frequency and amplitude (Table \ref{tab:feature}).
Dexterity features are extracted from the \textit{stay} duration, that is, the length of time the finger stays touching the screen, and the \textit{move} duration which is the interval of time between a finger release and the next finger press. The reaction features focus on the lag times of finger reactions (i.e. the time intervals between the stimulus appearing and the finger touch event).

\begin{table}%
\footnotesize
\centering
\caption{Brief description of features extracted for active tests}
\begin{threeparttable}
\centering
\begin{tabular}{p{1.5cm}|p{6cm}}
\toprule
\bfseries Feature & \bfseries Brief Description \\
\midrule
\bfseries Acceleration &  \textit{mean}: Mean\\
\bfseries Features\tnote{a} 	& \textit{std}: Standard deviation \\
	& \textit{Q1}: $25^{th}$ percentile \\
	& \textit{Q3}: $75^{th}$ percentile \\
	& \textit{IQR}: Inter-quartile range (IQR) $(Q3 - Q1)$ \\
	&  \textit{median}: Median\\
	&  \textit{mode}: Mode (the most frequent value)\\
	&  \textit{range}: Data range (max - min)\\
	&  \textit{skew}: Skewness\\
	&  \textit{kurt}: Kurtosis\\
	&  \textit{MSE}: Mean squared energy\\
	&  \textit{En}: Entropy\\
	&  \textit{MCR}: Mean cross rate\\
	&  \textit{DFC}: Dominant frequency component\\
	&  \textit{AMP}: Amplitude of DFC\\
	&  \textit{meanTKEO}: Instantaneous changes in energy due to body motion\tnote{b} \\
	&  \textit{AR1}: Autoregression coefficient at time lag 1\\
	&  \textit{DFA}: Detrended fluctuation analysis \cite{little2006nonlinear}\\
	&  \textit{XCORR}: Cross-correlation between two axes\\
	&  \textit{MI}: Mutural information between two axes\\
	&  \textit{xEn}: Cross-entropy between two axes\\
\midrule
\bfseries Voice &  \textit{Len}: Voice duration in seconds \\
\bfseries Features	& \textit{AMP}: Voice amplitude \\
	& \textit{F0}: Dominant voice frequency \\
	& AMP and F0 features include mean, standard deviation, DFA, and the coefficients of polynomial curve fitting with degree one and two respectively \\
\midrule
\bfseries Dexterity Features&  apply the same feature set (includes mean, standard deviation, Q1, Q3, IQR, median, mode, range, skew, kurt, MSE, En, meanTKEO, AR1, DFA) on two groups of tapping intervals: \\
	& \textit{STAY}: length of time finger stays touching the phone screen  \\
	& \textit{MOVE}: time interval between release of touch to the next touch event \\
\midrule
\bfseries Reaction Features &  apply the same feature set on the lags of finger reactions (i.e. the time intervals between the stimulus appearing and the finger touch event), including sum, mean, standard deviation, Q1, Q3, IQR, median, mode, range, skew, kurt, MSE, En, meanTKEO, DFA \\
\bottomrule
\end{tabular}
\begin{tablenotes}
\scriptsize
	\item[a]{Acceleration features are used for both balance and gait tests}
  \item[b]{TKEO stands for Teager-Kaiser energy operator \cite{kaiser1990simple}}
\end{tablenotes}
\end{threeparttable}
\label{tab:feature}
\end{table}

\subsubsection{Classification}
here we investigate whether the active tests can be used to detect dopaminergic medication response. 

We use a random forest classifier to generate a mapping from an active test instance to a discrimination of whether the instance represents off treatment (tests performed before medication) or on treatment (tests performed after medication has been taken). 
The random forest classifier \cite{breiman2001random} is an ensemble learning method for classification, regression and other machine learning tasks. This method fits many decision tree classifiers to randomly selected subsets of features and averages the predictions from each of these classifiers. We use a random forest classifier with 500 trees and the splitting criterion is based on Gini impurity\footnote{Gini impurity is a standard measure used in classification and regression trees (CART) \cite{breiman1984classification} to indicate the diversity a set of training targets. It reaches its minimum (zero) when all training cases in the node fall into a single target class.}.

Random forests can also be used to assess the relative importance of features in a classification problem. In our case, the importance of features reflects how strongly predictive they are of the effect of medication.


In addition, to compare against a ``null'' classifier, we use a na\"{\i}ve benchmark -- the random classifier\footnote{A random classifier predicts the unknown active test group by guesses based solely on the proportion of classes in the dataset, in this case, 50\% chance to be a baseline or a treatment instance.}. In theory this random classifier should achieve exactly 50\% performance accuracy.

We use 10-fold cross validation (CV) with 100 repetitions to estimate the out-of-sample generalization performance. For each run of 10-fold CV, the original dataset is randomly partitioned into 10 equal-sized subsets. Of the 10 subsets, a single subset is retained as the validation subset (10\% of the instances), and the remaining nine subsets used as the training data (90\% of the instances). The CV process is then repeated 10 times so that each of the 10 subsets is used exactly once as the validation set. We repeat this 10-fold CV process 100 times after permuting the original instance dataset uniformly at random. This provides a distribution of classification accuracies which allows estimation of the mean and standard deviation of the performance of the classifier for unseen datasets, hence controlling for overfitting of the classifier to the single available dataset.
Classification results are discussed in detail in Section \ref{sec:result_class}.

\section{Results}
In this study our aim is to develop a smartphone-based remote monitoring platform, to assess its feasibility for remote monitoring of symptoms in Parkinson's disease, and to demonstrate the value of data collected by detecting dopaminergic medication response. 
Here, we first illustrate the feasibility of this entirely remote PD monitoring by summarizing the characteristics of participants and the data collected in the worldwide PD study. Then, we discuss the discriminative performance on medication response detection and the ten most predictive features indicated by the random forest classifier.

\subsection{Data Collection Overview and Population Characteristics}
\label{sec:feasibility}
\begin{table}[t]
\centering
\caption{Baseline characteristics}
\scriptsize
\begin{tabular}{p{3cm}|c c}
	\toprule
		& PD (N = 121) & Control (N = 105) \\
	\textbf{Characteristic} & N (\%) or mean (SD) & N (\%) or mean (SD)\\
	\midrule
	\textbf{Demographic information} & & \\
	Gender (\% male) & 71 (59) & 56 (53) \\
	Age (years) & 57.6 (9.1) & 45.5 (15.5) \\
	Race (\% white) & 104 (86) & 86 (82) \\
	College graduate & 100 (83) & 76 (72) \\
	Previous participant in PD study? (\% yes) & 56 (46) & 11 (10) \\
	\midrule
	\textbf{Technology information} & & \\
	Duration of smartphone ownership ($>$1 year) & 111 (92) & 97 (93) \\
	Downloaded other apps previously? (\% yes) & 109 (90) & 98 (94) \\
	Search for health information using phone? (\% yes) & 111 (92) & 98 (94) \\
	\midrule
	\textbf{Clinical information} & & \\
	Care from PD specialist (\% yes) & 68 (56) & N/A \\
	Years since symptoms began & 5(19.8) & N/A \\
	Years since diagnosis & 5(4.7) & N/A \\
	Years on medication(s) & 5(4.7) & N/A \\
	\bottomrule
\end{tabular}
\label{tab:cha}
\end{table}
\begin{figure}[t]
	\centering
		\includegraphics[width=.45\textwidth]{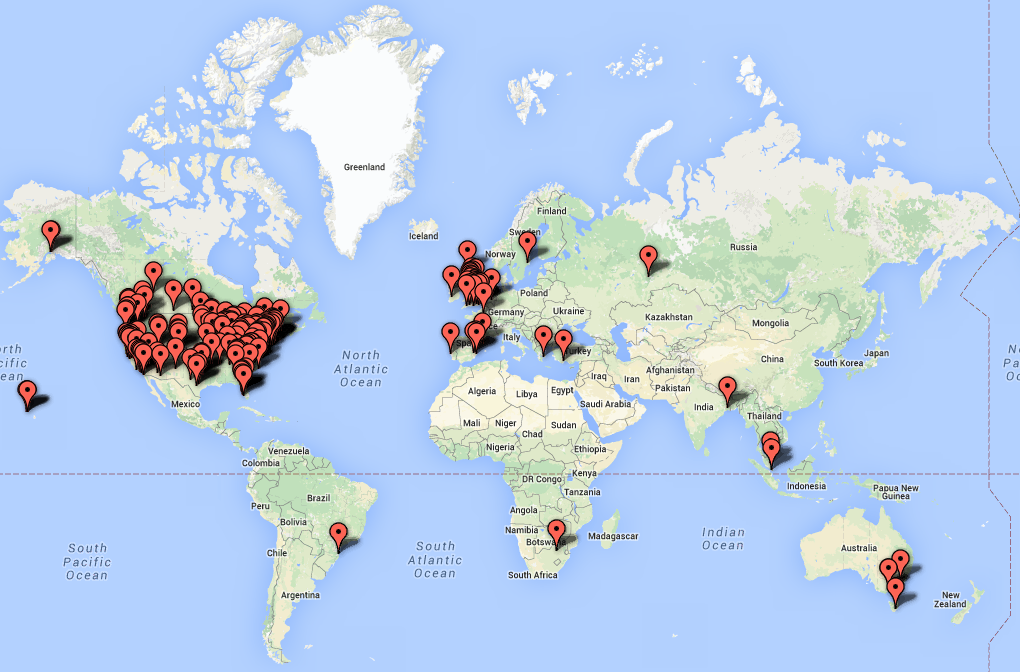}
	\caption{HopkinsPD enables worldwide remote monitoring. Each red pin shows the location of a participant.}
	\label{fig:global}
\end{figure}
\begin{figure}[t]
	\centering
		\includegraphics[width=0.49\textwidth]{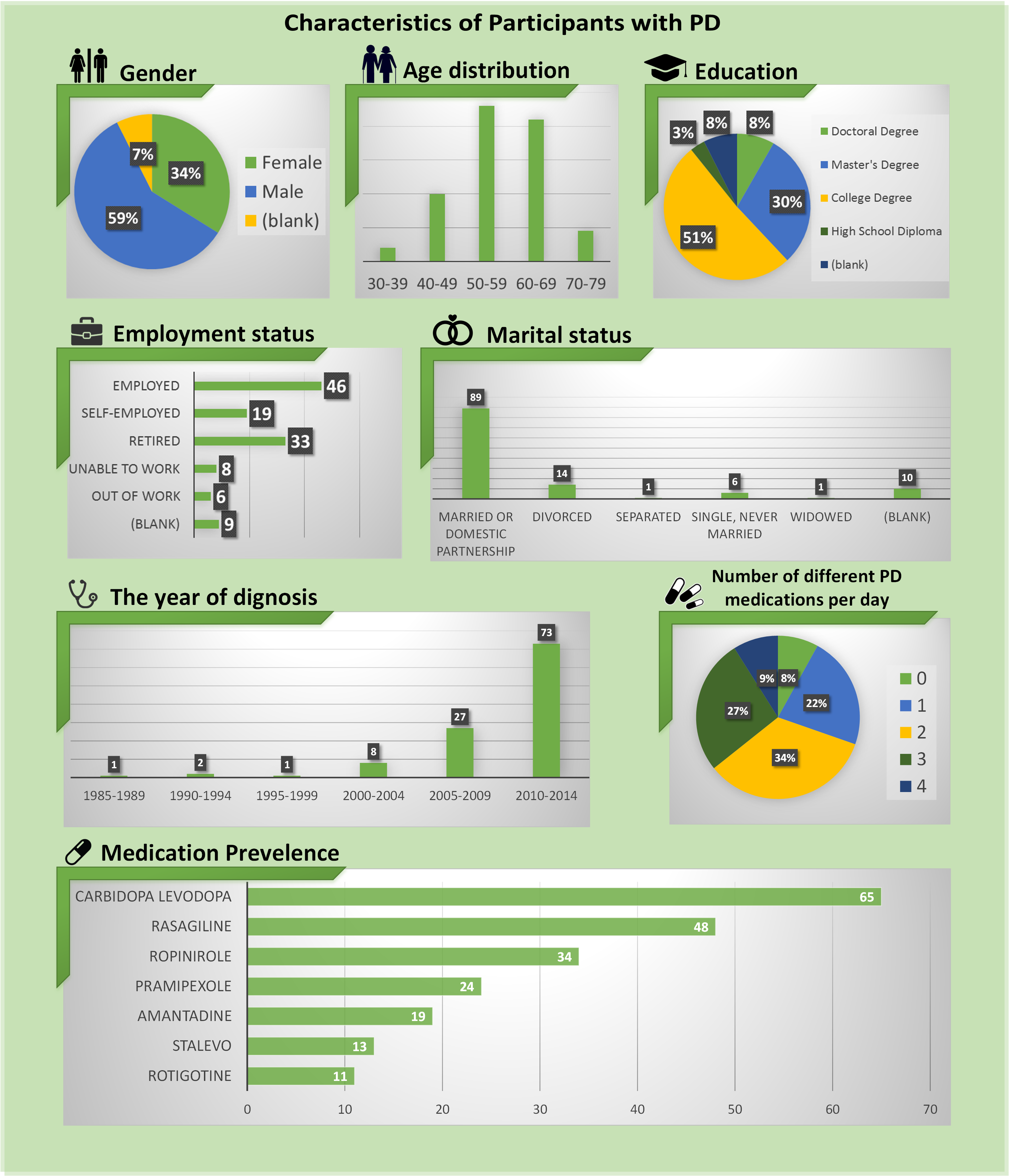}
	\caption{Detailed characteristics of PD participants}
	\label{fig:pd_cha}
\end{figure}

\begin{figure*}[ht]
\centering
\begin{subfigure}[b]{.5\textwidth}
  \centering
  \includegraphics[width=.9\linewidth]{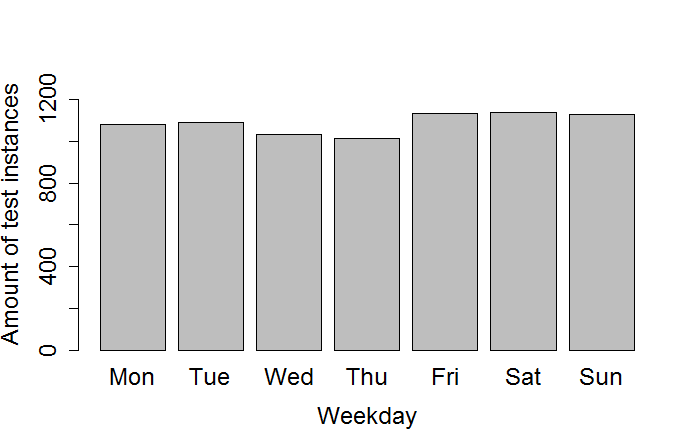}
  \caption{Instances of active tests collected in each day of the week}
  \label{fig:week_stats_ac}
\end{subfigure}%
\begin{subfigure}[b]{.5\textwidth}
  \centering
  \includegraphics[width=.9\linewidth]{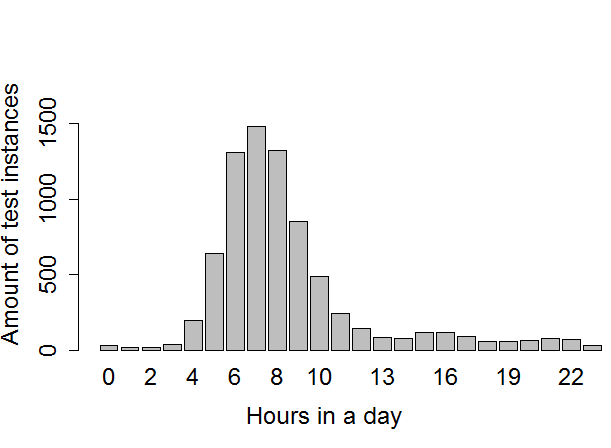}
  \caption{Instances of active tests collected in each hour of the day}
  \label{fig:day_stats_ac}
\end{subfigure}\\
\begin{subfigure}[b]{.5\textwidth}
  \centering
  \includegraphics[width=.9\linewidth]{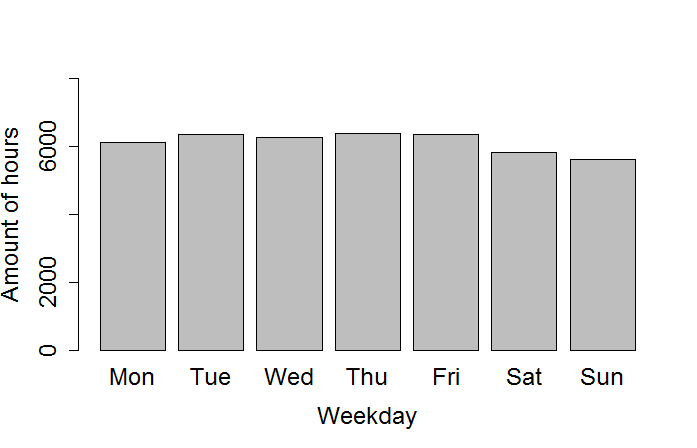}
  \caption{Hours of passive monitoring in each day of the week}
  \label{fig:week_stats}
\end{subfigure}%
\begin{subfigure}[b]{.5\textwidth}
  \centering
  \includegraphics[width=.9\linewidth]{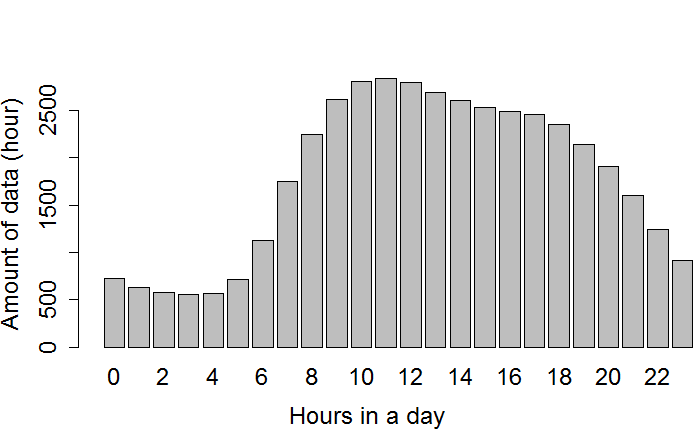}
  \caption{Hours of passive monitoring in each hour of the day}
  \label{fig:day_stats}
\end{subfigure}
\caption{Statistics of high frequency data collected from both active tests and passive monitoring}  
\label{fig:stats}
\end{figure*}

\begin{figure*}[ht]
\centering
\begin{subfigure}[b]{.48\textwidth}
  \centering
  \includegraphics[width=.98\linewidth]{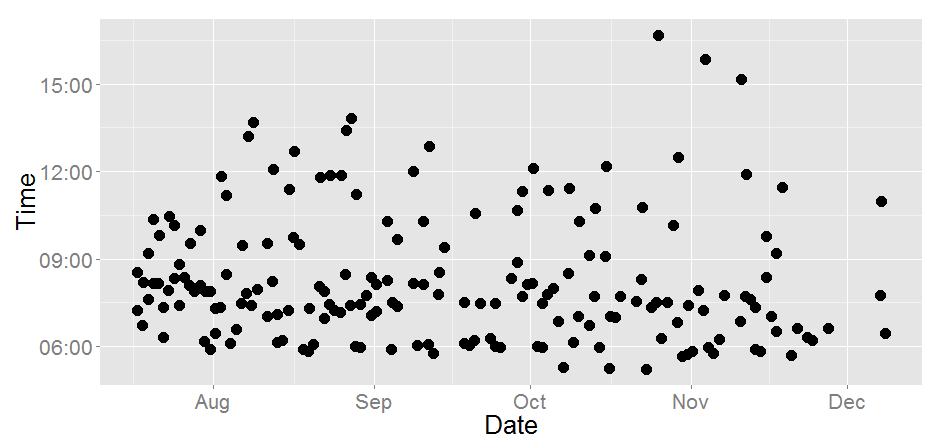}
  \caption{One hundred eighty five instances of active tests collected (black dots).}
  \label{fig:roch0065}
\end{subfigure}
\begin{subfigure}[b]{.48\textwidth}
  \centering
  \includegraphics[width=.98\linewidth]{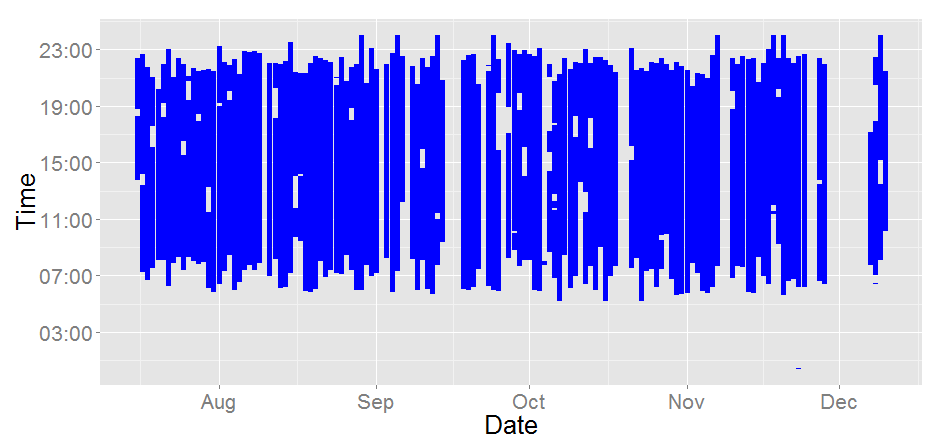}
  \caption{One hundred twenty six days of passive monitoring (each blue line segment represents one complete passive monitoring session)}
  \label{fig:roch0065}
\end{subfigure}%
\caption{High frequency data collected for a PD participant (roch0065) within six months}  
\label{fig:stats_ind}
\end{figure*}

Using the HopkinsPD platform, our study of remote PD monitoring launched in 2014. To the best of our knowledge, this is the first study to have collected continuous remote monitoring data at such a large scale in terms of participant size and participant duration. Any Android smartphone owner can join and contribute at no cost. We elaborate several advantages of our approach as follows:

For the first time, we have been able to run a PD-specific research study with worldwide participation. Table \ref{tab:cha} summarizes the characteristics of participants living with PD versus healthy controls. In our study, 226 individuals (121 PD and 105 controls) contributed data via HopkinsPD in the last half of 2014. As shown in Figure \ref{fig:global}, these participants come from many of the world's major population centers. 
Meanwhile, a considerable number of healthy controls were also enrolled as participants.  
The demographics of PD participants and healthy controls are similar, which is important because it will allow us to discover PD-specific distinctions between them in future. Moreover, the familiarity with smartphone usage among the PD participants is comparable to the healthy controls, which suggests that no special requirements are needed in PD-specific remote monitoring studies. This indicates that smartphone-based remote monitoring approaches which are feasible for the healthy population would also be feasible for PD research.
 
Moreover, with this number of PD participants we can gain insight into the population variability of PD; Figure \ref{fig:pd_cha} depicts the detailed characteristics of PD participants. The ages of PD participants range from the 30s to 70s, including young onset PD. Participants have varied education, employment and marital status as well. These participants are also in different stages of the disease, diagnosed from 1985 to 2014. Based on this, they would be on various medication regimes. For instance, more than ten types of anti-Parkinsonian drugs are used among them; two-thirds of them need to take more than one type of medication daily to manage their disease. This variety among PD participants helps to ensure that the results of any data analysis are as unbiased as possible.

Finally, we demonstrate high frequency data collection. As of publication, over 46,000 hours of unstructured streaming sensor data and 7,563 instances of structured tests of voice, gait, posture, and dexterity were collected within 6 months. This amount of data and the number of participants already far exceeds previous smartphone-based studies \cite{arora2013high} \cite{arora2015}. Figure \ref{fig:stats} summarizes all the data collected from both active and passive tests. Figures \ref{fig:week_stats_ac} and \ref{fig:week_stats} show the number of active test instances and the duration of passive monitoring by day of week respectively, showing weekly data volume collection is effectively uniform. Furthermore, Figures \ref{fig:day_stats_ac} and \ref{fig:day_stats} illustrate the number of active test instances and passive data collected by hour of day, respectively. They show most active tests being performed during the morning, and the passive monitoring primarily covering the daytime, particularly from 08:00 to 18:00. As an illustration of high frequency data collection, two timeline charts in Figure \ref{fig:roch0065} show all active tests and the periods of passive monitoring from a PD participant (ID: roch0065): this participant started the data collection on July 16, 2014 and recorded 185 instances of active tests and 126 days of passive monitoring. Despite pauses in data collection, which may be attributable to battery depletion, HopkinsPD was still able to collect passive data during most daytime hours.

Thus we demonstrate the feasibility of using a smartphone-based remote monitoring platform to collect high frequency data for large-scale PD research studies.

\subsection{Results of Medication Response Detection}
\label{sec:result_class}
7,653 instances of active tests were collected from both PD participants and healthy controls. To detect medication response for PD participants, we used all the active test instances from PD participants which can be paired into baseline and treatment, giving  4,388 instances in total, to train and evaluate the random forest classifier.

To quantify the performance of the random forest classifier in detecting medication response, we calculate three commonly used \textit{performance measures}:
\begin{itemize}
	\item Sensitivity (true positive rate) -- proportion of treatment instances correctly identified.
	\item Specificity (true negative rate) -- proportion of baseline instances correctly identified.
	\item Accuracy -- proportion of both treatment and baseline instances correctly identified.
\end{itemize}

From Table \ref{tab:acc} we can see that the accuracy of the random forest classifier is considerably higher than that of random guessing, which indicates that medication response is detectable by using active tests ($p < 0.001$, two-sided Kolmogorov-Smirnov test). Based on these results, we can reject the null hypothesis that random forests have no discriminative power in detecting medication response.

\begin{figure*}[htb]
	\centering
		\includegraphics[width=\linewidth]{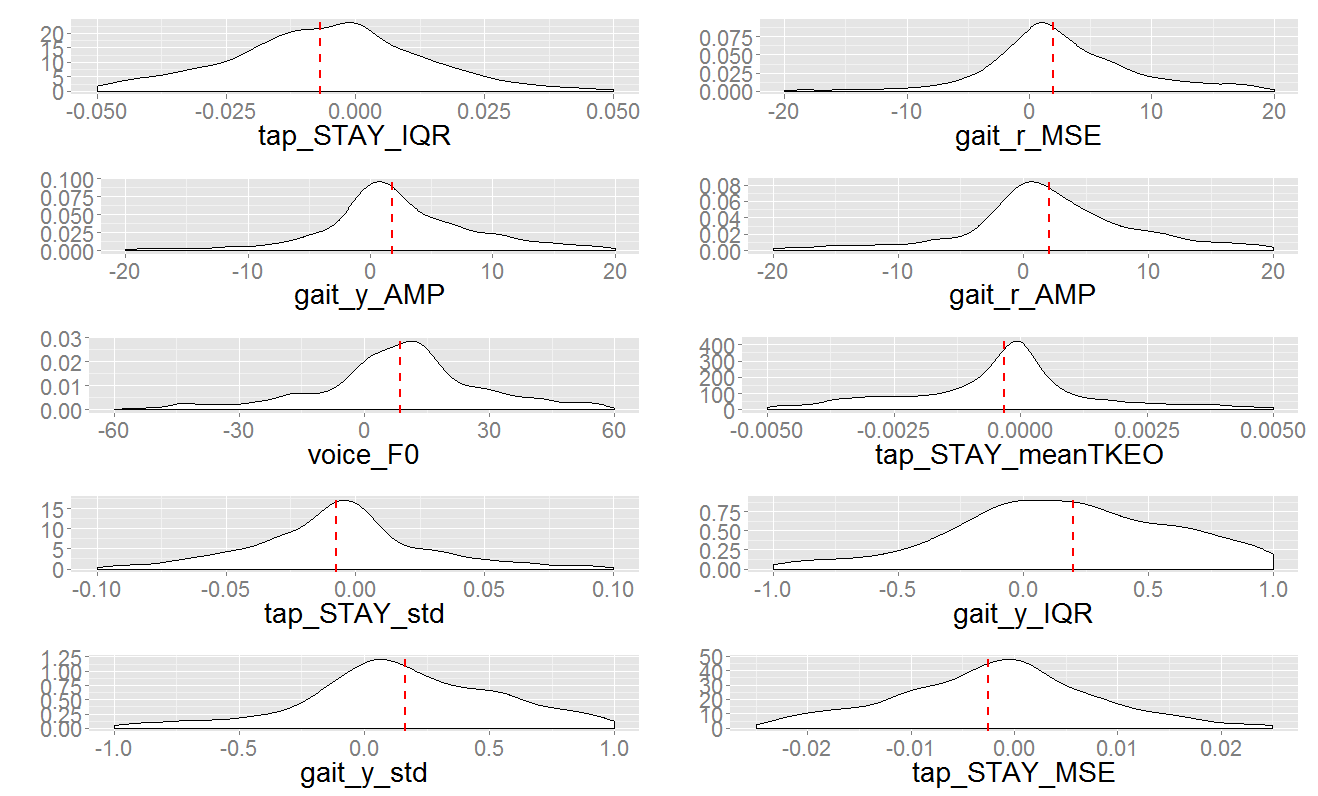}
	\caption{Probabilty density of feature differences from treatment to baseline among all participants (red dashed lines at median differences). The shift of the median differences from zero indicates the improvement on these features after taking medication.}
	\label{fig:diff_all}
\end{figure*}

The random forest classifier also indicates the importance of features in creating the classification trees. The ten most important features are described in Table \ref{tab:top_features}, and the density plots of the feature differences between baseline and treatment are depicted in Figure \ref{fig:diff_all}. Specifically, the following dexterity, voice, and gait features are the most useful in predicting response to dopaminergic medication:
\begin{itemize}
	\item Improved tapping rhythm -- The decrease of inter-quartile range, standard deviation, mean squared energy and the instantaneous energy changes of the finger pressing intervals suggests that finger tapping movements become faster and more stable after taking medication. This phenomenon indicates that medication relieves the PD symptoms directly related to hand movements, e.g. bradykinesia.
	\item Increased voice pitch -- An increase in vocal pitch after medication suggests that medication may relieve specific dysphonias such as monopitch.
	\item Improved gait -- The increased inter-quartile range, standard deviation, and amplitude of the acceleration signals during walking indicates that PD participants walk more vigorously after taking medication. More specifically, the changes of these features on axis y are more significant than axes x and z. Since the y axis points to the ground during the gait test, it indicates PD participants can lift their feet higher off the floor after taking medication.
\end{itemize}

\begin{figure}[htb]
	\centering
		\includegraphics[width=0.45\textwidth]{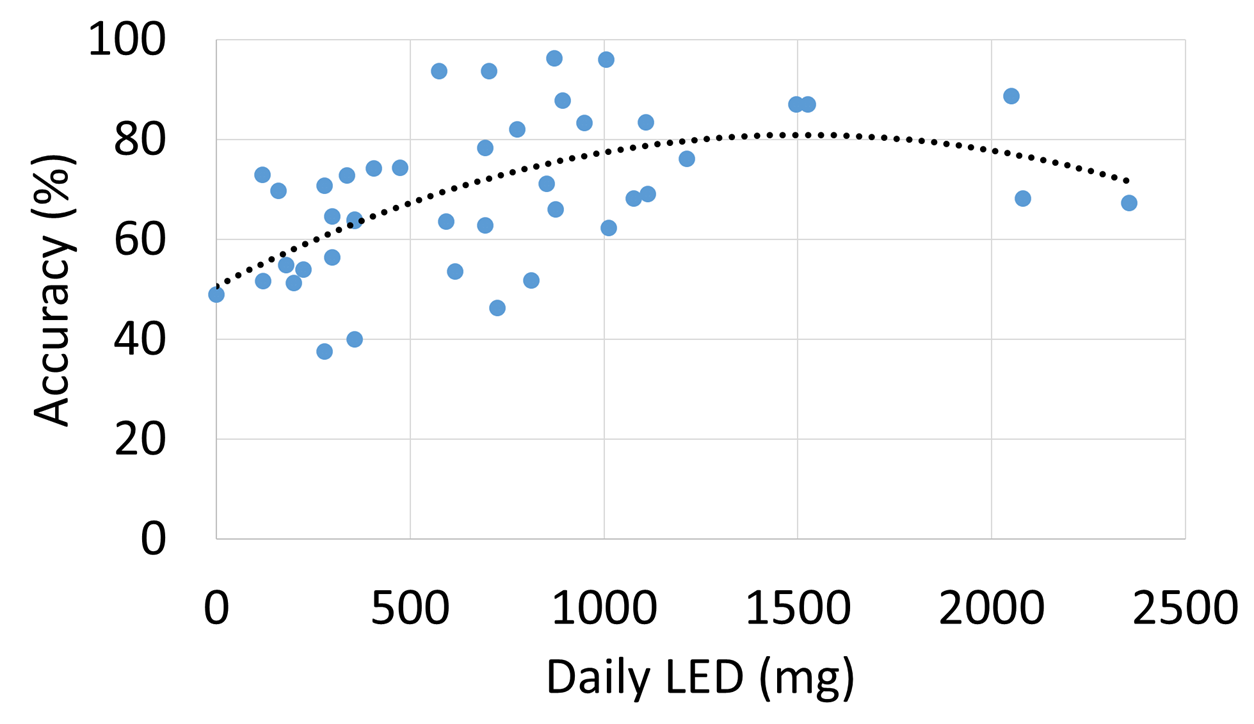}
	\caption{The relation between accuracy and daily LED}
	\label{fig:ledvsacc}
\end{figure}

We also find that the accuracy of medication response detection varies substantially with total daily levodopa equivalent dose (LED) \cite{tomlinson2010systematic}. Here, the daily LED is computed via a standardized formula from daily drug regimes, reported in the pre-study survey \cite{tomlinson2010systematic}. Daily LED provides a useful tool to express dose intensity of different anti-Parkinsonian drug regimes on a single scale, and we assume that these regimes do not change within six months. As shown in Figure \ref{fig:ledvsacc}, each point indicates the average accuracy of medication response detection for each individual\footnote{Here we only show the individuals with more than twenty active test instances} versus the individual's daily LED. The dotted line is the quadratic polynomial regression line. The medication response can be detected more accurately with LED between 500 and 2000 mg than for low dose (less than 500 mg) or high dose (larger than 2000 mg). This is consistent with the fact that individuals with higher dosage of medications are more likely to have (motor) fluctuations in their disease and thus detectable changes in response to treatment.

\begin{figure*}[htb]
\centering
\begin{subfigure}[b]{.5\textwidth}
  \centering
  \includegraphics[width=.99\linewidth]{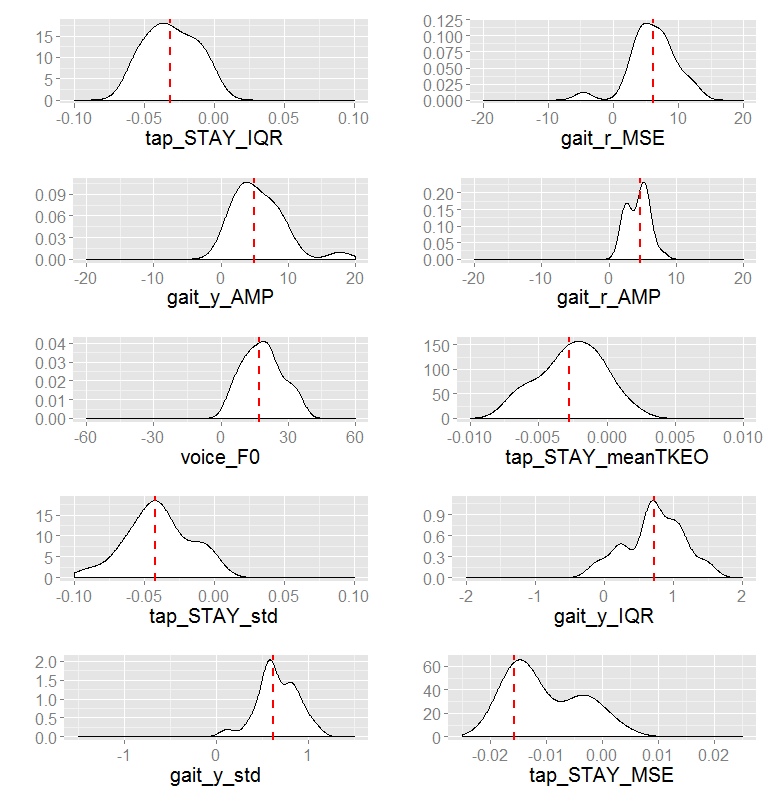}
  \caption{User: roch0064, daily LED = 872 mg}
  \label{fig:roch0064}
\end{subfigure}%
\begin{subfigure}[b]{.5\textwidth}
  \centering
  \includegraphics[width=.99\linewidth]{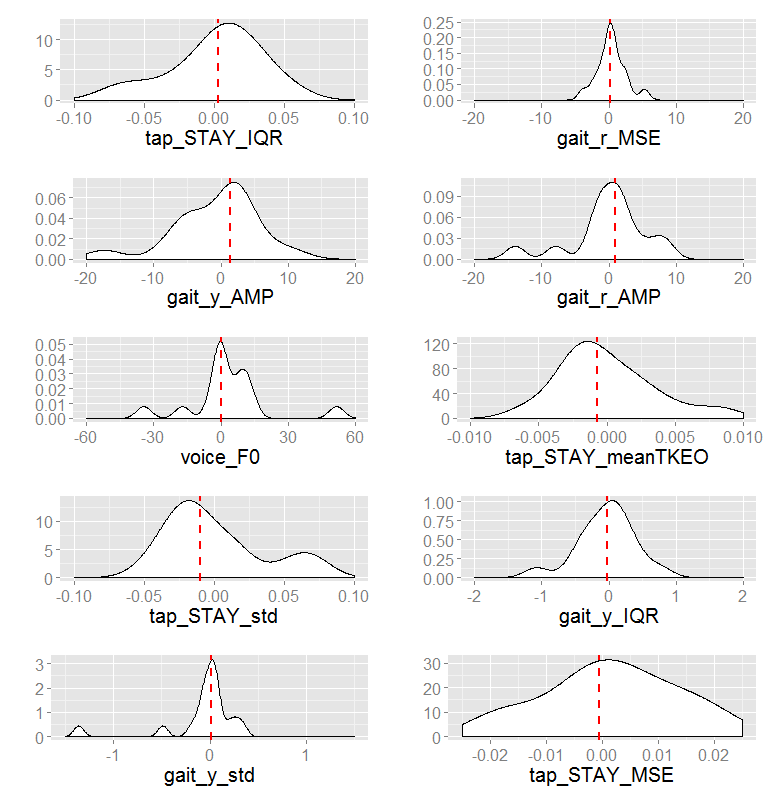}
  \caption{User: roch0359, daily LED = 120 mg}
  \label{fig:roch0359}
\end{subfigure}%
\caption{Probability density plots of the feature differences from treatment to baseline from two PD participants (red dashed lines at median differences). The participant roch0064, taking a medium LED (872 mg), exhibits a distinct improvement on the ten features while the improvement is not obvious on roch0359, an individual on low LED (120 mg).}  
\label{fig:diff_ind}
\end{figure*}

To further understand this, Figure \ref{fig:diff_ind} compares the different medication responses between two individuals with different LEDs. Participant roch0064, taking a medium LED (872 mg), exhibits a distinct improvement on the ten features, while the improvement is not obvious on roch0359, an individual on low LED (120 mg). We speculate that for individuals with low LED, the dose is too small to detect a clinically important difference between baseline and treatment; however, for individuals on the largest LED, usually with advanced symptom severity, the medication ``wears off'' in between doses, but high dosage also induces side effects such as dyskinesias. Further investigation is required to validate these conjectures.

\begin{table}[t]
\caption{Performance measures of the methods used for medication response detection}
\centering
\begin{threeparttable}
\centering
\begin{tabular}{llll}
	\toprule
	\textbf{Method}	& \textbf{Sensitivity} &	\textbf{Specificity} & \textbf{Accuracy} \\
	\midrule
	\textbf{Random Forest} &  $69.3 \pm 0.5$ & $72.7 \pm 0.1$ & $71.0 \pm 0.4$ \\
	\textbf{Random Classifier} & $49.0 \pm 0.1$ & $50.9 \pm 0.9$ & $50.0 \pm 0.6$ \\
\bottomrule
\end{tabular}
\begin{tablenotes}
	\item Note: the results are reported in the form average $\pm$ standard deviation in percentages (\%). 
	\item Sensitivity = TP/(TP+TN)
	\item Specificity = TN/(TN+FP)
	\item Accuracy = (TP+TN)/(TP+TN+FP+FN)\\
	Here TP, TN, FP, and FN stand for true positive, true negative, false positive, and false negative, respectively.
\end{tablenotes}
\end{threeparttable}
\label{tab:acc}
\end{table}

\begin{table}[t]
\footnotesize
\caption{The top ten features selected by the random forest classifier}
\centering
\begin{threeparttable}
\centering
\begin{tabular}{llp{3.6cm}}
	\toprule
	\textbf{Feature ID}	& \textbf{Test} & \textbf{Description}  \\
	\midrule
	{tap\_STAY\_IQR} & Dexterity & inter-quartile range of finger pressing intervals \\
	{gait\_y\_AMP} & Gait & the amplitude of the dominant frequency on axis y \\
	{voice\_F0} & Voice & the dominant voice frequency \\
	{tap\_STAY\_std} & Dexterity & standard deviation of finger pressing intervals \\
	{gait\_y\_std} & Gait & standard deviation of axis y \\
	{gait\_r\_MSE} & Gait & mean squared energy of the radial distances \\
  {gait\_r\_AMP} & Gait & the amplitude of the dominant frequency of the radial distances \\		
	{tap\_STAY\_meanTKEO} & Dexterity & mean TKEO of finger pressing intervals \\
  {gait\_y\_IQR} & Gait & inter-quartile range of axis y \\
	{tap\_STAY\_MSE} & Dexterity & mean squared energy of finger pressing intervals \\
\bottomrule
\end{tabular}
\end{threeparttable}
\label{tab:top_features}
\end{table}

\section{Discussion}


In this study, we demonstrate that various objective measurements of PD symptoms can be collected on a single smartphone-based platform. Compared with consumer wearable devices and specialized medical equipment, smartphones are far more pervasive (wearable devices are used by only less than 10\% of online adults, according to a recent survey \cite{gwi}). Moreover, our smartphone-based platform is a more flexible and comprehensive toolkit compared to previous studies based on a single wearable device. We believe that the smartphone is the optimal ``central hub'' for remote monitoring of symptoms in PD. Wearable devices like a wristband or a smart watch could be useful accessories for certain important purposes, e.g. sleep monitoring. It may also be that, in the future, a smart watch with sufficient capability could entirely replace the smartphone. Regardless, it is possible to integrate wearable devices into HopkinsPD. Data streams from future wearable devices would be another kind of passive test, and these data can be synchronized to the phone via Bluetooth and securely uploaded to the server using the existing framework. This would allow raw sensor data collected from wearable devices to be accessed by PD researchers.

Importantly, in our PD study, hundreds of individuals with PD around the world, downloaded this PD-specific smartphone application. This has provided tens of thousands of hours of objective data on PD. We have shown that, these data coming from individuals in their home environment, can be used to detect medication response, namely whether the medication improves dexterity, voice, and gait among as expected. We have also observed that many participants do not improve as much as others. The possible reasons include suboptimal medication regimes or wearing-off and motor complications, e.g. levodopa-induced dyskinesia. Further understanding and verification of individualized daily motor fluctuations will be key to closing the loop for integrated PD symptom monitoring and intervention (Figure \ref{fig:vision}).

The current study is completely observational with no interventional aspects. Future efforts could include the ability to provide near real-time feedback to participants on their performance to enable them to have more control over their condition. The functionality to complete more detailed surveys such as the MDS-UPDRS, and the incorporation of additional passive monitoring may also be valuable.


\section{Conclusion}
Current clinical monitoring of PD is low-frequency and based on periodic clinic visits. In contrast, we propose HopkinsPD, a novel mobile smartphone-based platform to monitor PD symptoms frequently and remotely. 
Our feasibility study has shown that this approach enabled remote PD monitoring on an unprecedented worldwide scale with very high frequency data collection. By using the controlled tests carried out using only the built-in smartphone sensors, initial experiments using a set of features extracted from the sensor data and random forest classification, is able to detect medication response with 71.0($\pm$0.4)\% accuracy. Given the feasibility of using smartphone-based monitoring platform to collect high frequency clinical data and these preliminary results on medication response detection, future studies could devise novel models to track both the fluctuations in symptoms and medication response over time in order to assess disease progression, medication adherence, and, finally, help clinicians make decisions on appropriate therapeutic interventions informed by objective, near real-time, data.

\section*{Acknowledgment}

The authors would like to thank all the volunteers that participated in
this study; this paper would not have been possible
without them. We also wish to thank Fahd Baig for his feedback on testing of the smartphone software created in this study.

\ifCLASSOPTIONcaptionsoff
  \newpage
\fi


\bibliographystyle{IEEEtran}
\bibliography{pdbib}

\section*{Supplemental Section: HopkinsPD Implementation}
\label{sec:impl}

\begin{figure}[t]
\centering
\begin{minipage}{.15\textwidth}
  \begin{subfigure}{\linewidth}
  \centering
  \includegraphics[width=.9\linewidth]{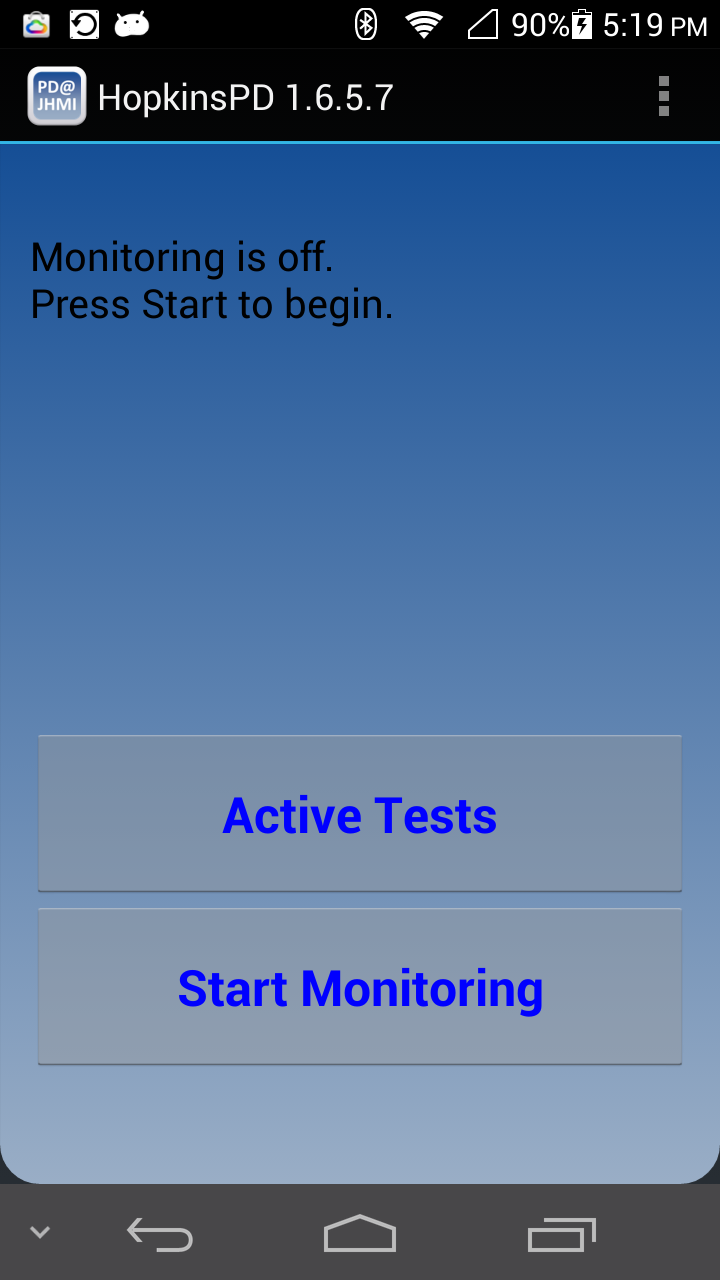}
  \caption{Primary interface}
  \label{fig:mainactivity}
\end{subfigure}%
\end{minipage}%
\begin{minipage}{.15\textwidth}
  \begin{subfigure}{\linewidth}
  \centering
  \includegraphics[width=.9\linewidth]{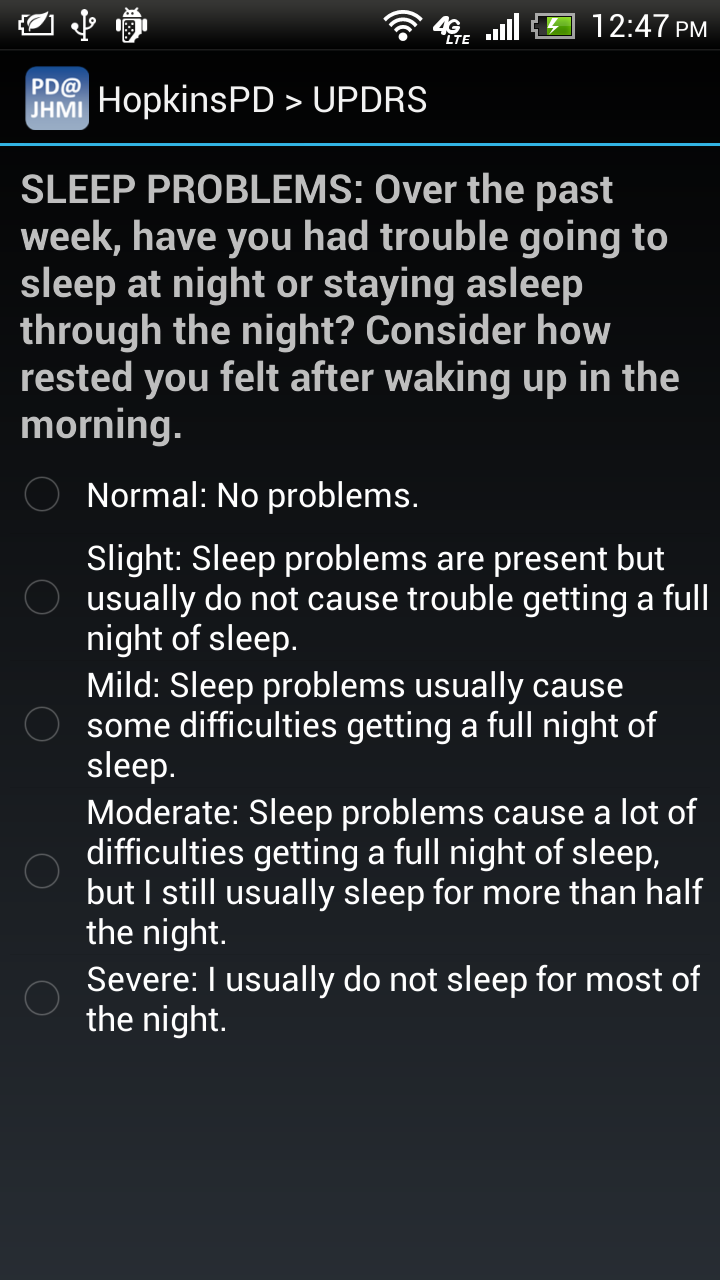}
  \caption{Self-report}
  \label{fig:selfreport}
\end{subfigure}%
\end{minipage}%
\begin{minipage}{.15\textwidth}
\centering
  \begin{subfigure}{\linewidth}\centering
    \includegraphics[width=.9\linewidth]{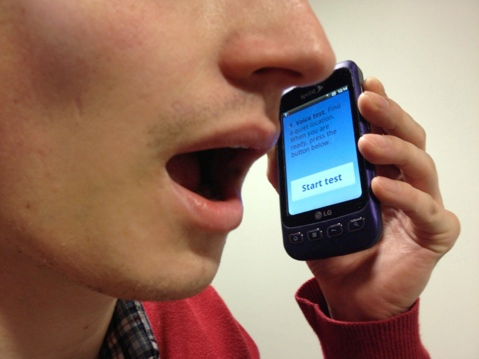}
  \label{fig:voicetest}
\end{subfigure}\\[1ex]
\begin{subfigure}{\linewidth}
  \includegraphics[width=.9\linewidth]{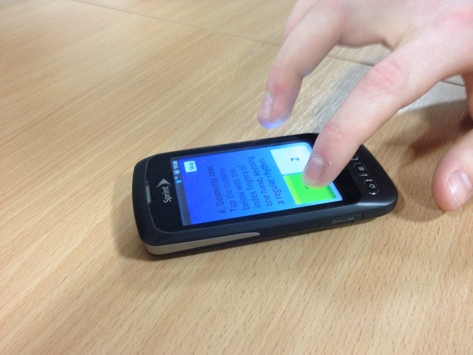}\\
  \caption{Active tests}
  \label{fig:dexteritytest}
\end{subfigure}%
\end{minipage}%
\caption{HopkinsPD mobile application.}  
\label{fig:app}
\end{figure}

To support all PD-specific tests on the smartphone application (shown in Figure \ref{fig:app}) and to facilitate study management for PD researchers, we focus on two functional aspects of the system:

\subsection{Web-based Project Management and Visualization}
\begin{figure}[t]
\centering
\begin{subfigure}[b]{.5\textwidth}
  \centering
  \includegraphics[width=.98\linewidth]{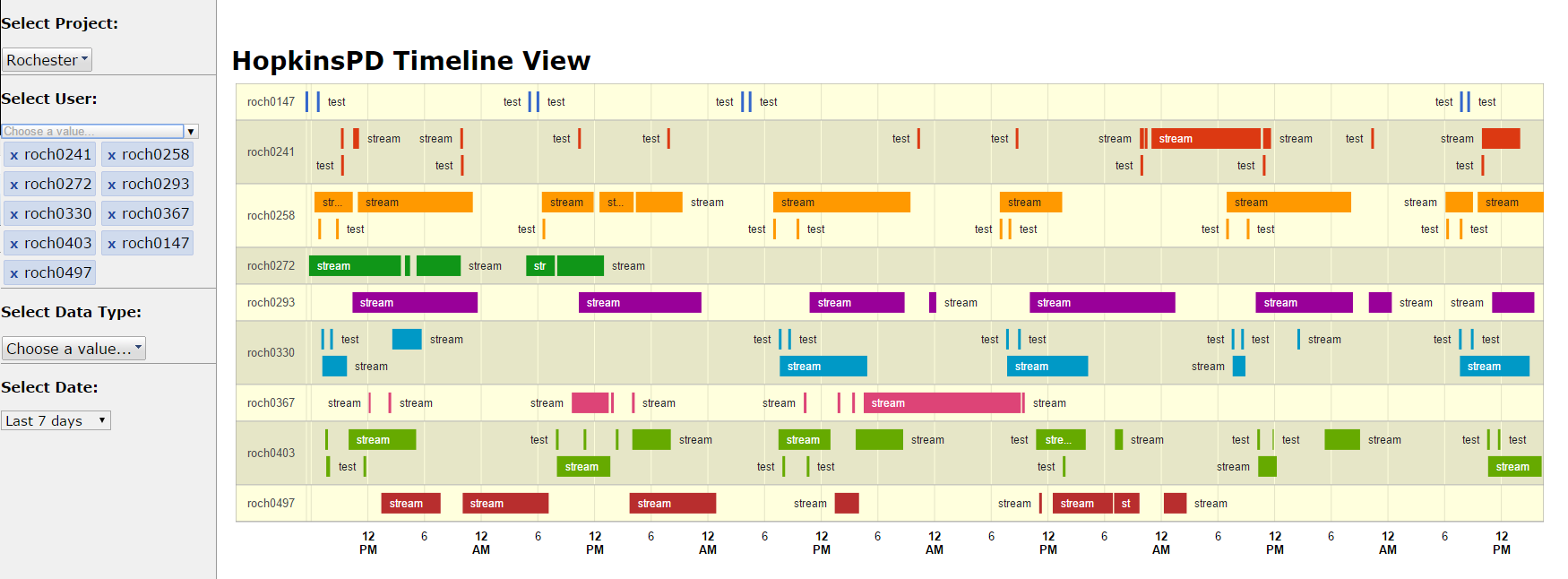}
  \caption{Timeline: user monitoring}
  \label{fig:timeline}
\end{subfigure}%
\hfil
\begin{subfigure}[b]{.5\textwidth}
  \centering
  \includegraphics[width=.98\linewidth]{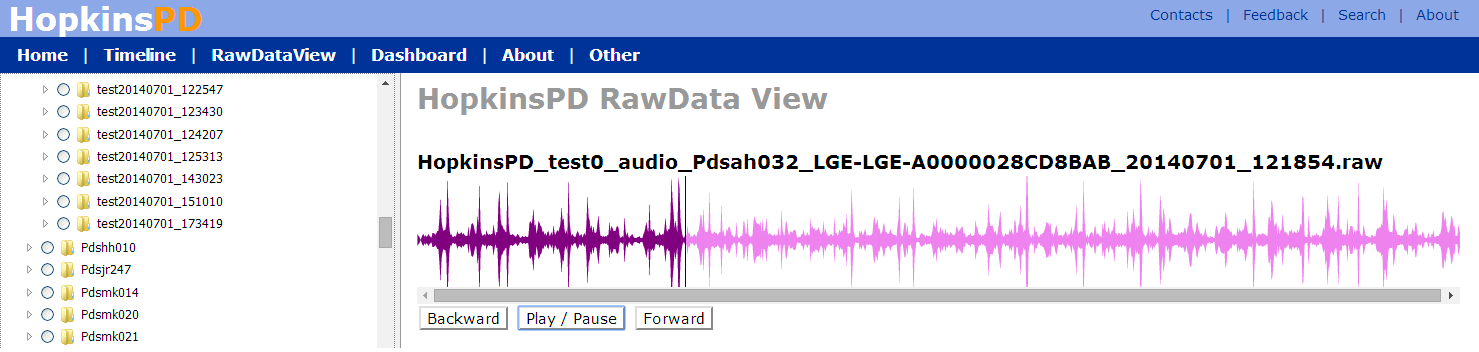}
  \caption{An example of test view: voice view}
  \label{fig:voiceview}
\end{subfigure}%
\hfil
\begin{subfigure}[b]{.5\textwidth}
  \centering
  \includegraphics[width=.98\linewidth]{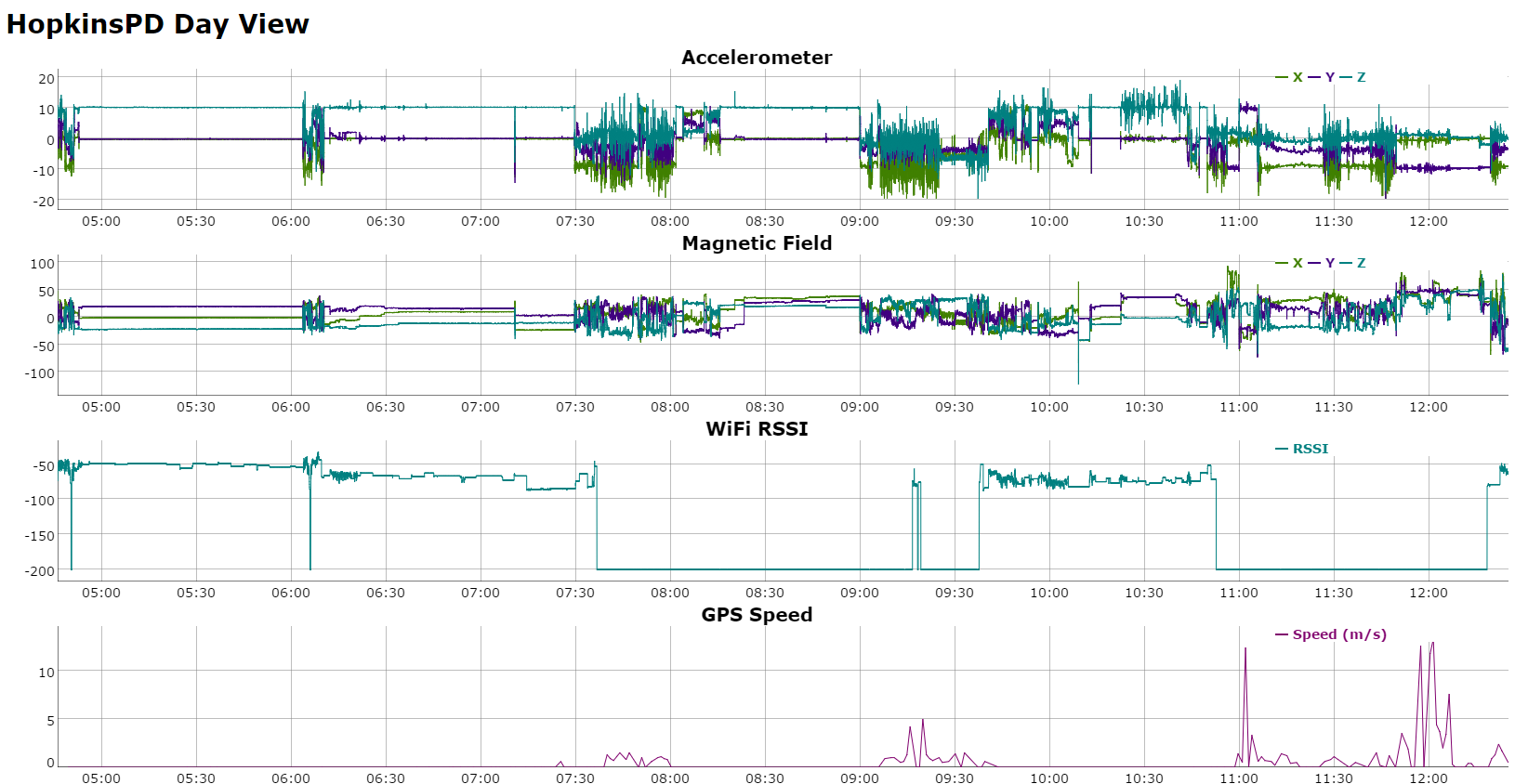}
  \caption{Part of the day view}
  \label{fig:voicetest}
\end{subfigure}%
\caption{HopkinsPD web-based visualization.}  
\label{fig:web}
\end{figure}

The HopkinsPD web front-end is designed to make PD studies manageable for researchers. It provides:
\begin{itemize}
	\item Detailed project configuration. HopkinsPD allows researchers to customize the application before launching. In particular, researchers can configure which active tests are enabled, which sensor data to collect during passive monitoring, and which questionnaires to activate. To bring this to light, they can turn on or turn off GPS data collection and configure the sampling frequency, e.g., once per minute or once per hour. The project configuration is scripted using a simple XML file. 
	\item Usage monitoring. A \textit{timeline} view is provided for researchers to track data collection progress remotely, as well as user participation in near real-time. An interactive interface is provided in this view so that researchers can specify a range of subjects and time. An example is shown in Figure \ref{fig:timeline}.
	\item Multi-level visualization. To explore and visualize the multidimensional data from both active and passive tests, there are two different views provided: a) a test \textit{detailed view} displaying detailed sensor data plots with zooming. For instance, researchers can visualize the three dimensional acceleration sensor data collected from the accelerometer when the subject is performing a gait test, or play the sound recorded during the voice test; b) a \textit{day view} summarizing daily passive tests on a dashboard consisting of various charts representing time-series of movement sensor data (e.g., acceleration and compass), location (GPS coordinates on a map), users' interaction with the phone (e.g. app usage and phone calls), and resource consumption (e.g. battery level). This allows researchers to monitor users' daily activities. Examples of data visualization are shown in Figure \ref{fig:web}.
\end{itemize}
%


\subsection{HIPAA-Compliant Security}
To meet HIPAA requirements, the system must not broadcast identifiable patient data and must guarantee the authenticity of the data it captures.  
HopkinsPD is designed to be HIPAA-compliant to maintain the integrity of Protected Health Information (PHI) for all participants, which is necessary in that even though personal information such as names are excluded from the platform, PHI could be inferred from data collected from smartphones. Taking GPS as an example, coordinates collected on the phone may expose the participants' home or work address. Therefore, the backend of HopkinsPD has implements the following:
\begin{itemize}
	\item All data is immediately encrypted after collection on the phone; secondly, all data uploaded to the server and all results generated from that data are encrypted and stored on a managed protected server with restricted access. 
	\item Data transmission security. HopkinsPD implements a unified upload manager which uploads data collected via the mobile frontend. This includes HTTPS-based encrypted upload, error handling, and retry mechanisms. 
	\item Access control and authentication. The account management and access control are support authorized individual data access in concurrent deployment settings. A HopkinsPD account can be created by an administrator. Users are authenticated based on their account credentials. The access control in HopkinsPD is designed at a project level so that for an account authorized to access Project A, data collected from participants in Project A alone are accessible to this account. This design isolates projects from one another, thus allowing multiple studies to be securely managed and deployed on the same backend. 
\end{itemize}

The above functionalities greatly assist researchers in running remote PD studies and provide a secure data streaming service to protect the data collection process.
\end{document}